\documentclass[twocolumn,floatfix,showpacs,
prl,superscriptaddress]{revtex4-1}
\usepackage{graphicx}% Include figure files
\usepackage{float}
\usepackage{dcolumn} % Align table columns on decimal point
\usepackage{bm}
\usepackage{epsfig}
\usepackage{natbib}
\usepackage{amsmath}
\usepackage{mathtools}
\usepackage{listings}
\usepackage{longtable}
\usepackage{csquotes}
\usepackage{gensymb}
\usepackage{xcolor}
\usepackage{hyperref}
\usepackage{comment}
\hypersetup{
     colorlinks   = true,
     citecolor    = blue
}

\newcommand{\out}[1]{}

\begin{document} 

\title{Electronic structure of CuTeO$_4$ and its relationship to cuprates}

\author{Antia S. Botana}
\affiliation{Materials Science Division, Argonne National Laboratory, Argonne, Illinois 60439, USA}
\author{Michael R. Norman}
\affiliation{Materials Science Division, Argonne National Laboratory, Argonne, Illinois 60439, USA}

\begin{abstract}

Based on first principles calculations, the electronic structure of CuTeO$_4$ is discussed in the context of superconducting cuprates.
Despite some significant crystallographic differences, we find that CuTeO$_4$ is similar to these cuprates, exhibiting a quasi two dimensional 
electronic structure that involves hybridized Cu-$d$ and O-$p$ states in the vicinity of the Fermi level, along with an antiferromagnetic 
insulating ground state.  Hole doping this material by substituting Te$^{6+}$ with Sb$^{5+}$ would be of significant interest.

\end{abstract}

\maketitle

High-temperature superconductivity in cuprates is one of the most intriguing emergent phenomena in strongly correlated electron systems \cite{cuprates}. Some common features of cuprates include their layered crystal structures, proximity to a magnetic insulating state, and hybridization of Cu-$d$ and O-$p$ orbitals. The investigation of materials with features in common with these is key to establish the importance of these various features for superconductivity and to provide guidance in the search for new
high temperature superconductors. 

In spite of an intense search for new cuprates, copper tellurates  \cite{review} have not been explored for potential superconductivity. A number have been reported in
regards to their crystal structures.  Those where Cu is 2+ and Te is 6+ include CuTeO$_4$ \cite{falck}, Cu$_3$TeO$_6$ \cite{cu3teo6}, Sr$_2$CuTeO$_6$ \cite{sr2cuteo6} (and its Ba variant \cite{ba2cuteo6}), Ag$_4$CuTeO$_6$, \cite{ag4cuteo6}, and Na$_2$Cu$_2$TeO$_6$ \cite{na2cu2teo6}. Several also exist
where Te can be in a 4+ state (CuTe$_2$O$_5$, PbCuTe$_2$O$_6$, SrCuTe$_2$O$_7$) that we do not discuss here.  Some of these materials have
been investigated for their interesting magnetic properties, but in all these cases, their crystal structures do not resemble those of cuprates.
Ag$_4$CuTeO$_6$ and Na$_2$Cu$_2$TeO$_6$ have Cu-O-Te-O-Cu chains, Cu$_3$TeO$_6$ has a 3D array of Cu hexagons,
and though Sr$_2$CuTeO$_6$ exhibits a square planar net, the Cu ions are connected by a super-superexchange
pathway (Cu-O-Te-O-Cu) as in the chain materials.  The exception is CuTeO$_4$, which is the focus of this Letter.

\begin{figure}
\center
\includegraphics[width=\columnwidth,draft=false]{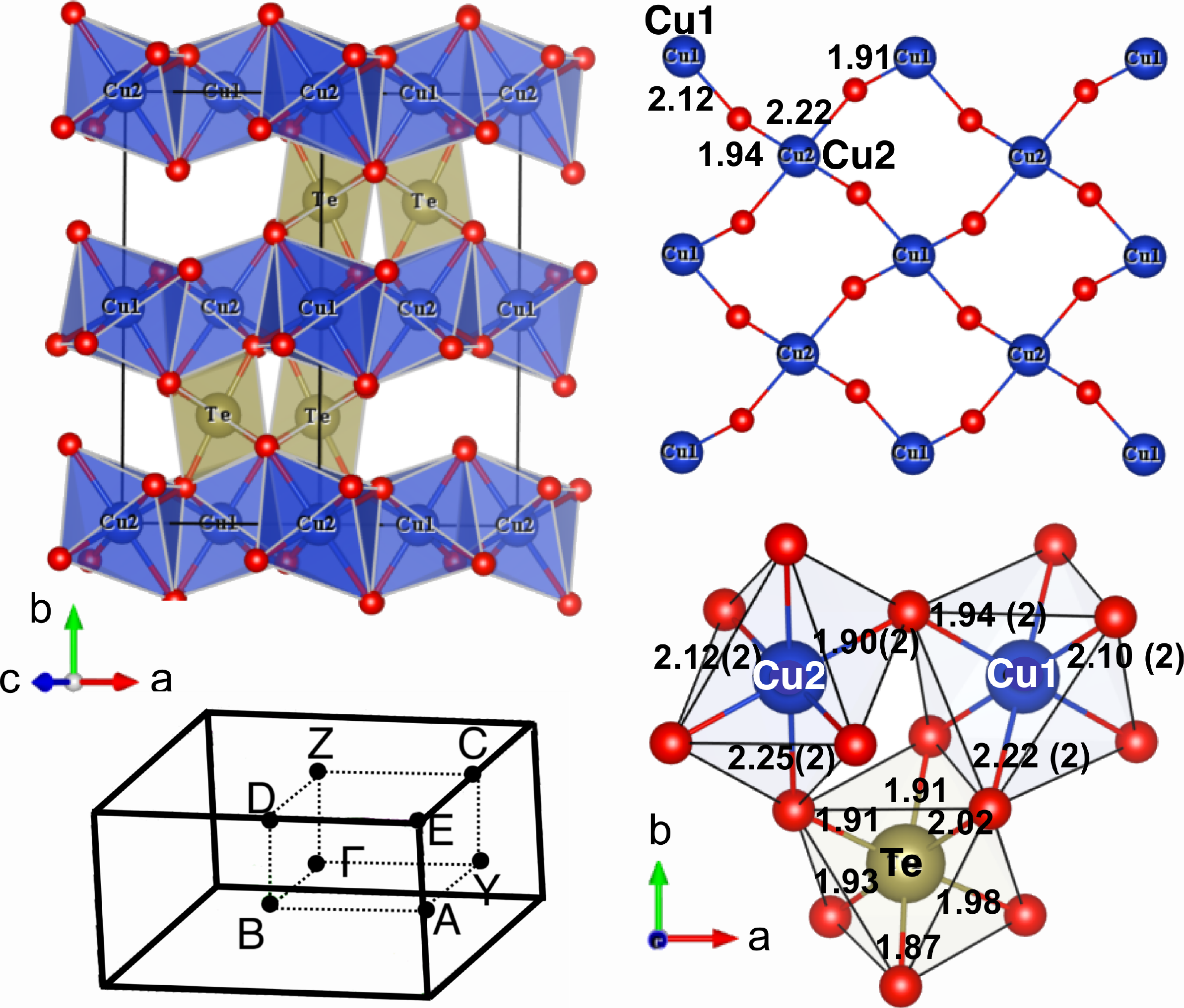}\\
\caption{Crystal structure of CuTeO$_4$ illustrating the buckled CuO$_2$ planes (upper right) that result from coupling of CuO$_6$ octahedra to TeO$_6$ octahedra
(upper left). Also shown is the corresponding Brillouin zone with high symmetry points marked (lower left), as well as the local environment of the Cu and Te ions
(lower right)  Note that the orthorhombic zone has axes that are rotated 45$^{\circ}$ relative to a tetragonal cuprate.
}\label{struct}
\end{figure}

Even though its crystal structure was reported almost forty years ago \cite{falck}, the electronic and magnetic properties of CuTeO$_4$ have not been explored.  This may be
due to the difficulty of growing the material (Cu$_3$TeO$_6$ tends to form instead \cite{gospodinov}), and the fact that the crystals are disordered, probably due to stacking faults.  On the
other hand, unlike the above mentioned materials,  CuTeO$_4$ exhibits CuO$_2$ planes, though they are highly buckled (Fig.~\ref{struct}).
This occurs because, unlike in most cuprates, the copper ions are in a distorted octahedral environment with Cu-O bond lengths varying by less than 20\%. Moreover,
the planes are coupled by TeO$_6$ octahedra which share oxygen atoms with the CuO$_6$ octahedra, driving the strong buckling with the `planar' Cu-O bonds titled from 24.1$^\circ$ to 28.4$^\circ$ out of the plane.
 Regardless, our electronic structure calculations show considerable similarities of this material to the superconducting cuprates, though with some interesting differences.

Electronic structure calculations were performed within density functional theory \cite{dft,dft_2} using the all-electron, full potential code WIEN2k \cite{wien2k} based on an augmented plane wave plus local orbitals (APW+lo) basis set \cite{sjo}. We have used the Perdew-Burke-Ernzerhof version of the generalized gradient approximation (PBE-GGA) for the exchange-correlation potential \cite{gga}. Calculations including spin-orbit coupling are shown in Ref.~\onlinecite{suppl}.
The parameters used were $R_{mt}K_{max}$= 7.0, with muffin-tin radii of 1.93 for Cu, 1.89 for Te, and 1.63 for O (a.u.). Brillouin zone integrations were
done with a 10$\times$16$\times$5 mesh. We performed calculations both with the experimental structural parameters taken from Falck {\it et al.}~\cite{falck} and with fully relaxed atomic positions.

Though the crystals are disordered, Falck {\it et al.}~\cite{falck} report a monoclinic structure with P2$_1/n$ symmetry. CuTeO$_4$ has 4 formula units per unit cell (Z=4) containing two inequivalent Cu sites, four inequivalent oxygens and one Te site. As mentioned above, the Cu ions exhibit a highly distorted octahedral environment with 2 short, 2 medium and 2 long Cu-O distances ranging from 1.91 to 2.26 \AA~for Cu1 and 1.94 to 2.22 \AA~for Cu2 (Fig.~\ref{struct} and Table \ref{table1}). The CuO$_2$ planes are not as regular as in a typical cuprate, with alternating Cu1 and Cu2 ions, and Cu-O-Cu bond angles of 122.5$^{\circ}$ or 126.1$^{\circ}$ along nearly orthogonal directions (81.1$^{\circ}$ and 98.9$^{\circ}$). It should be noted that the Cu2-O long bonds lie within the buckled planes (Fig.~\ref{struct}).  The TeO$_6$ octahedra are less distorted, with Te-O distances ranging from 1.87 to 2.02 \AA. The oxygens in the CuO$_2$ planes are shared by the Te ions causing the buckling of these planes.
The structural relaxation performed within GGA-PBE (Table \ref{table1}) tends to reduce the octahedral distortion by nearly balancing the short and medium Cu-O distances. This tendency points towards the possibility that the reported short-medium-long pattern of the Cu-O bonds can be due to a dynamic Jahn-Teller effect \cite{burns}
since the structure was determined at room temperature.  Detailed measurements of the structure and its atomic displacement parameters as a function of temperature could shed light on this.

\begin{table}
\caption{Cu-O and Te-O distances in the relaxed and unrelaxed structures. Note the different environments for the two inequivalent Cu sites.  L, M, and S are used to specify the long, medium and short Cu-O bonds.}\label{table1}
\begin{center}
\begin{tabular}{ c c c c c c c c  }
\hline
\hline
 &unrelaxed&  &  & & & &    \\
\hline
     &       &       O1 & (L) & O3 & (M) & O4 & (S) \\
&Cu1&   2.26 & 2.26 & 2.12 & 2.12 & 1.91 & 1.91 \\
\hline
& &  O4 & (L) & O2 & (M) & O3 & (S) \\
&Cu2 &    2.22 & 2.22 & 2.11 & 2.11& 1.94 & 1.94  \\
\hline
& &  O4 & O3 & O1 & O2 & O1 & O2 \\
&Te & 1.87 & 1.91 & 1.91 & 1.93 & 1.98 & 2.02\\
\hline
\hline
 &relaxed&  &  & & & &    \\
 \hline
      &       &       O1 & O1 & O3 & O3 & O4 & O4 \\
&Cu1 &    2.22 & 2.22 & 2.01 & 2.01& 1.98 & 1.98  \\
\hline
& &  O4 & O4 & O2 & O2 & O3 & O3 \\
&Cu2&   2.22 & 2.22 & 2.03 & 2.03 & 1.93 & 1.93 \\
\hline
& &  O4 & O3 & O1 & O2 & O1 & O2 \\
&Te & 1.90 & 1.93 & 1.94 & 1.97 & 1.99 & 2.02\\
\hline
\end{tabular}
\end{center} 
\end{table}

\begin{figure}
\center
\includegraphics[width=\columnwidth,draft=false]{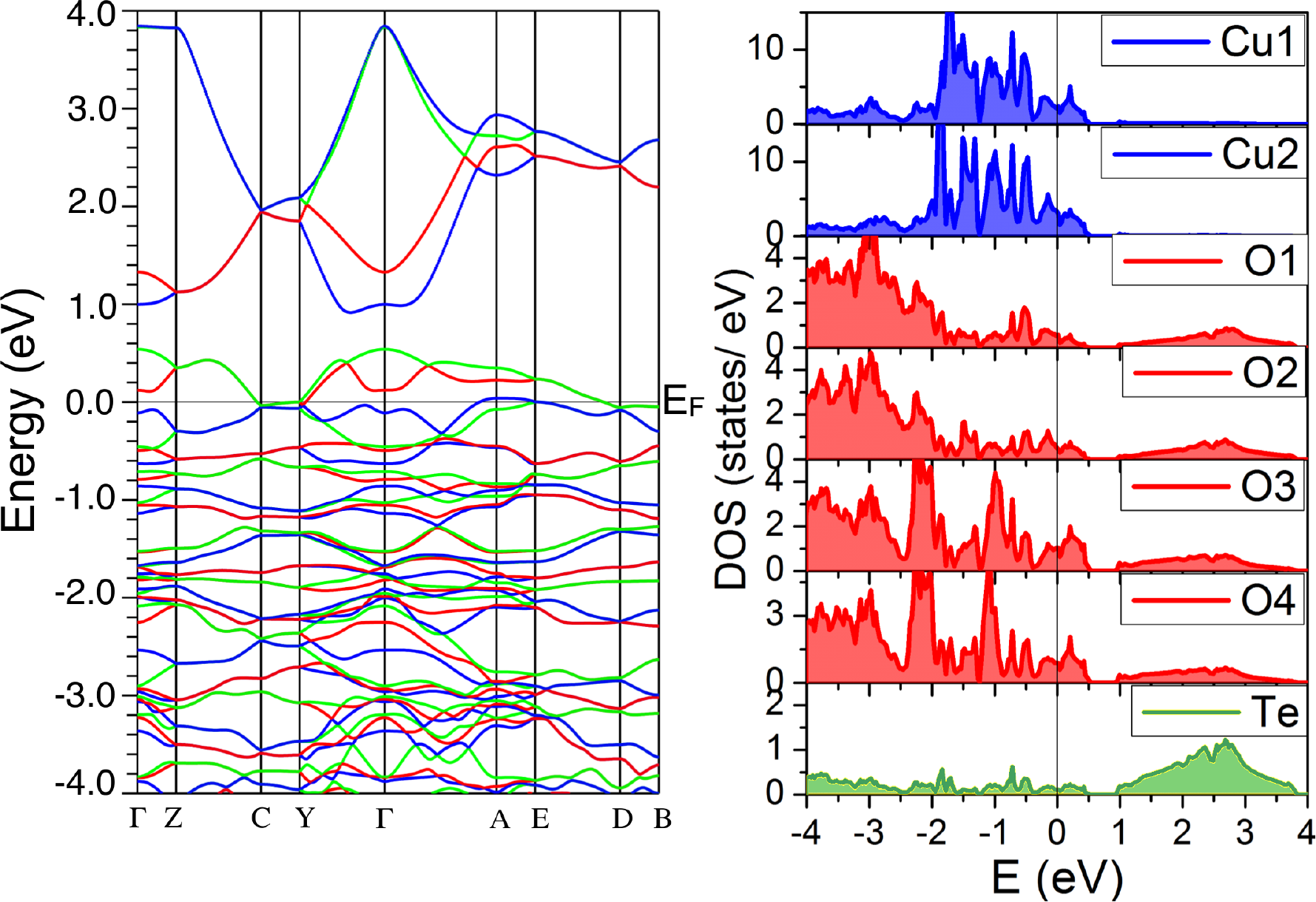}
\caption{Left: Paramagnetic band structure of CuTeO$_4$ for the experimental structural data. Four Cu $d$ - O $p$ hybridized bands cross the Fermi level. Right: Atom resolved density of states for the Cu, O and Te ions.}\label{bs_pm}
\end{figure}

Assuming a completely ionic character, Cu would be in a 2+ state (d$^9$).
The paramagnetic band structure, which is metallic because
of the odd number of electrons per cell, is shown in Fig.~\ref{bs_pm} along several high symmetry directions in the Brillouin zone as well as the corresponding density of states (DOS).

The Cu-$3d$ and O-$2p$ states form a complex of hybridized valence bands with a total bandwidth of $\sim$ 9 eV. In spite of the complexity of the valence-band manifold, the band structure around $E_F$ is rather simple with four bands crossing the Fermi level. With the $t_{2g}$ states being fully occupied for the Cu atoms, these bands have contributions from both Cu1 and Cu2 and are an admixture of $d_{x^2-y^2}$ and $d_{z^2}$-like states (due to the 2-2-2 nature of the Cu-O
bond lengths) with strong O-$2p$ hybridization.
The strong hybridization between these states is evident given the similar shapes of the O and Cu DOS in the vicinity of the Fermi level more evident for O3 and O4 (the `planar' oxygens) than for O1 and O2.  The wide bands appearing above 2 eV correspond mainly to Te-$5s$ states hybridized with O-$2p$ states. There is no significant contribution of Te states around the Fermi level, pointing towards the quasi-two dimensionality of the electronic structure.

\begin{figure}
\center
\includegraphics[width=\columnwidth,draft=false]{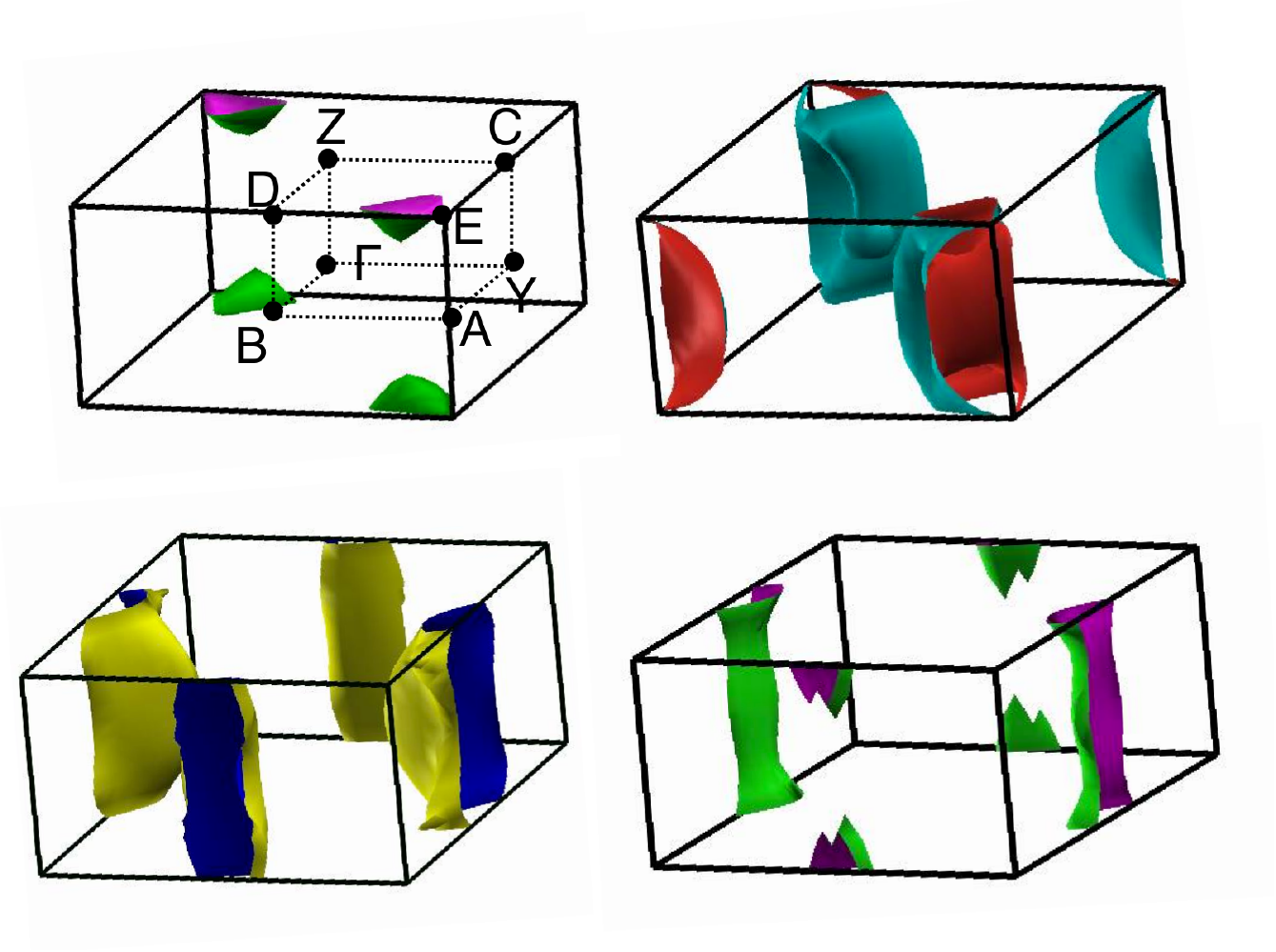}\\
\caption{The Fermi surfaces for CuTeO$_4$ from the four bands in the paramagnetic state (top: hole-like bands; bottom: electron-like bands).}\label{fs}
\end{figure}

Turning to the band structure, the dispersion perpendicular to the CuO$_2$ planes is small along certain zone directions ($C$-$Y$) but not as small along
others ($\Gamma$-$Z$).  The latter is characteristic of zone folding (Z=4), with the absence of splitting at $Z$ due to band sticking from
the screw axis.
Folding is also seen along in-plane directions, which is clearest along $Z$-$C$, with the splitting being very small at $C$ (0.02 eV).
More complicated
behavior is seen along $\Gamma$-$Y$ for two of the four bands due to their differing coupling along the $b$ direction orthogonal to the planes.
In this context, note that Cu1 and Cu2 atoms are related by (1/2,0,1/2) and (0,1/2,0) translations,
and again all bands in the $Z$-$C$-$E$-$D$ zone face are doubly degenerate due to the screw axis.

The symmetries mentioned above are also reflected in the Fermi surfaces. For a typical cuprate, one finds warped cylinders. This is reminiscent of CuTeO$_4$ where one of the hole surfaces is nearly cylindrical, centered at the zone corners ($A$-$E$), and two of the electron ones also nearly cylindrical, centered instead at the midline of the zone faces ($A$-$E$-$D$-$B$ and $A$-$E$-$C$-$Y$), with the electron and hole bands due to zone folding as in orthorhombic La$_2$CuO$_4$ \cite{pickett}. The remaining Fermi surfaces show a strong orthorhombic distortion, coupled with a more 3D-like behavior.

\begin{figure}
\center
\includegraphics[width=\columnwidth,draft=false]{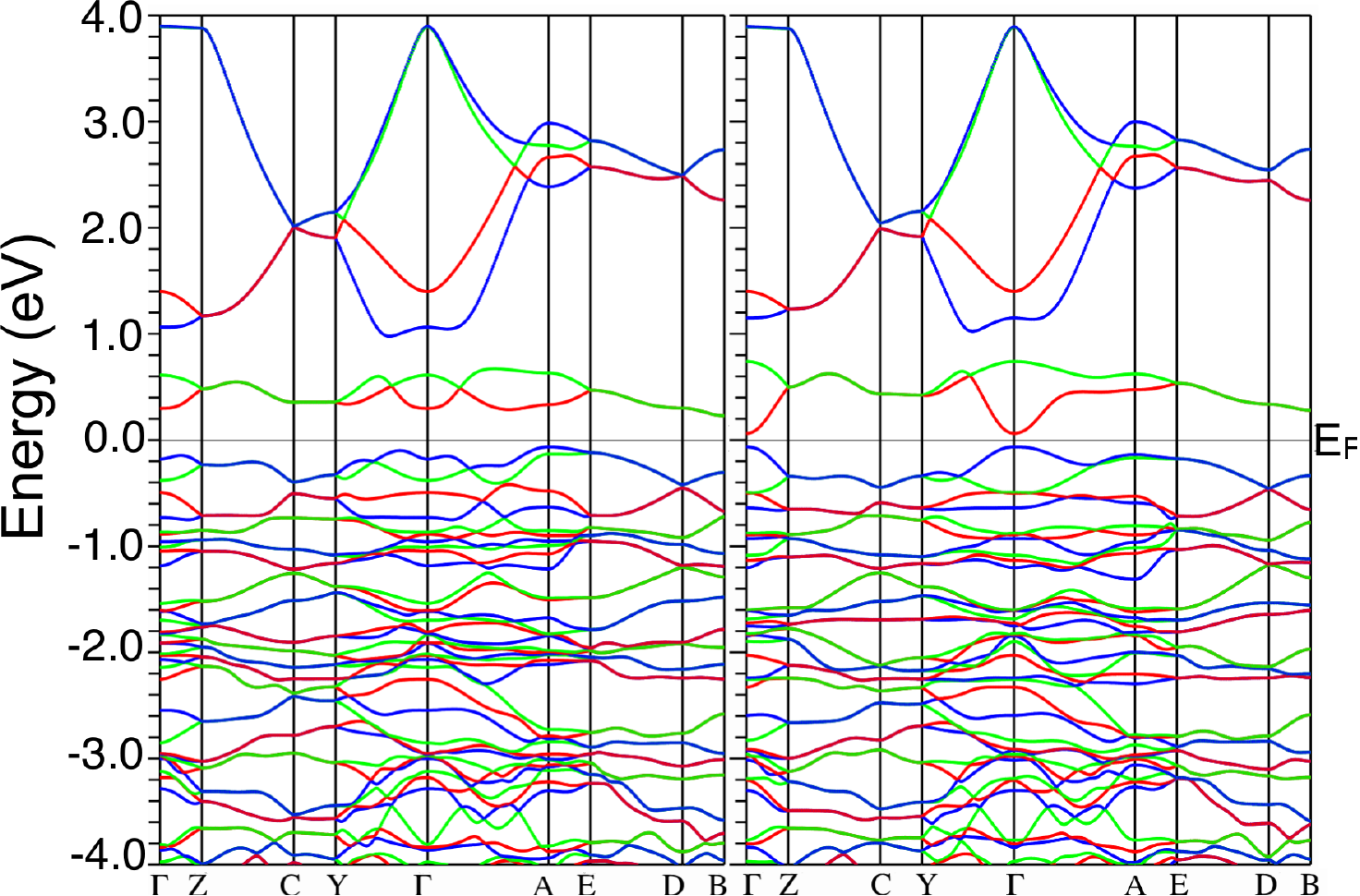}\\
\caption{Band structure of  antiferromagnetic (AFM) CuTeO$_4$ for the experimental structural data for the two spin channels (left: down; right: up). A small gap of 0.13 eV is apparent.}\label{bs}
\end{figure}

\begin{figure} 
\center
\includegraphics[width=\columnwidth,draft=false]{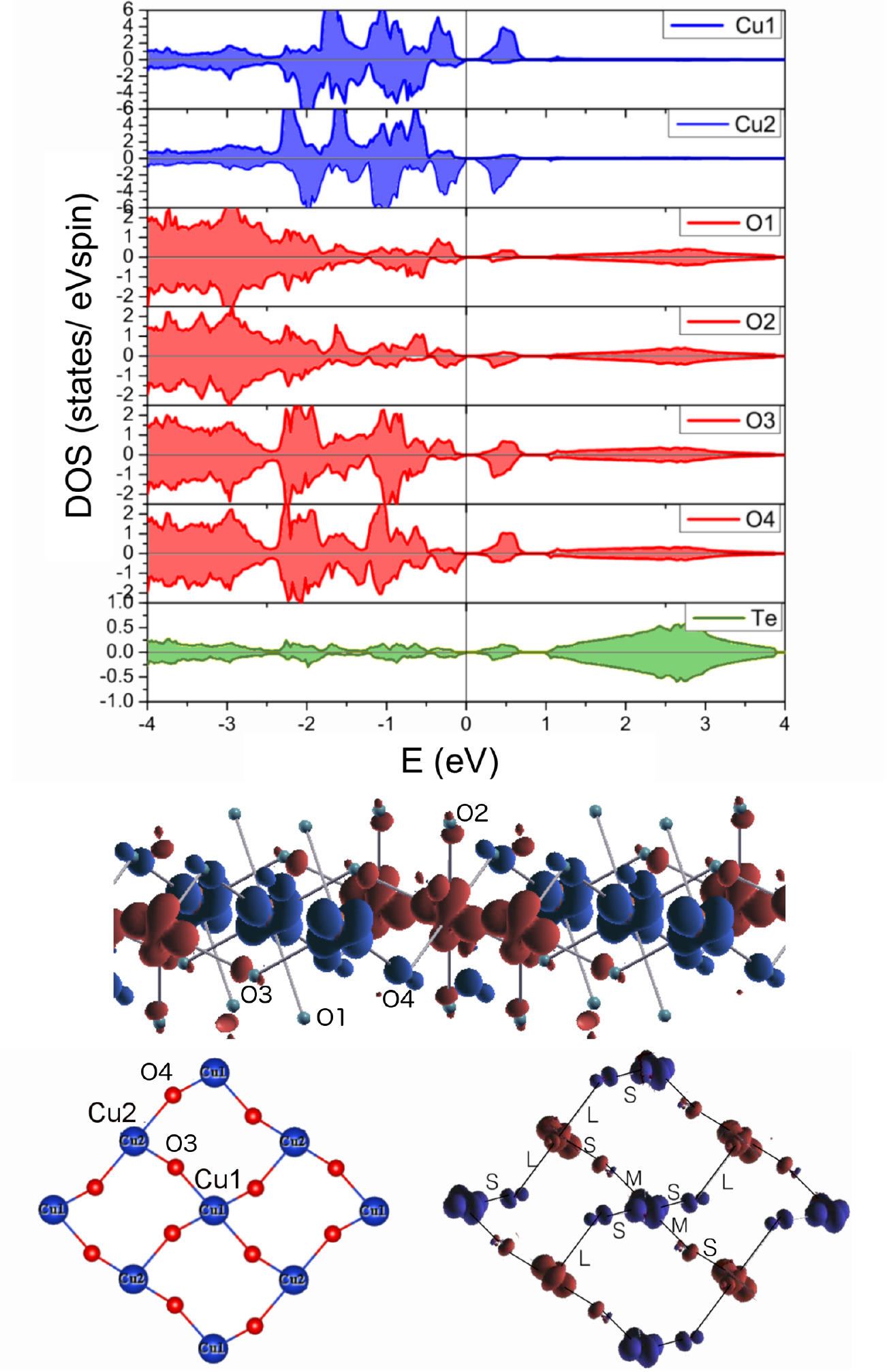}
\caption{Top panel: Atom resolved DOS for AFM CuTeO$_4$, showing a small gap within the hybridized Cu-$d$, O-$p$ manifold of states.  The
positive and negative values on the y-axis are for spin-up
and spin-down, respectively. Middle panel: Three-dimensional plot of the spin density in one of the Cu-O planes in the AFM state, with an isosurface at 0.1 e$^-$/\AA$^3$. Different colors represent the spin-up (spin-down) density. Bottom panel: view of the spin density in the CuO$_2$ planes, with short, medium, and long
Cu-O bonds indicated (S, M, L).}\label{dos}
\end{figure}

Given the magnetic character of the $d^9$ ion, one might expect the paramagnetic state to be susceptible to magnetic order.  We indeed find this to be the case, with the energy of the antiferromagnetic (AFM) state (with oppositely oriented spins on the Cu1 and Cu2 ions) being lower than that for the paramagnetic state by 34 meV/formula unit, and more stable than ferromagnetic ordering by 35 meV/formula unit.
The corresponding band structure for the AFM state is shown in Fig.~\ref{bs} and the atom-resolved DOS in Fig.~\ref{dos}. A small gap of $\sim$ 0.13 eV opens up due to the magnetic ordering, even without a Coulomb $U$, yielding a S=1/2 AFM insulator. The gap becomes larger once a $U$ is included as shown in Ref.~\onlinecite{suppl}. The gap is formed between minority-spin states only given that the crystal field splitting between the two $e_g$ orbitals is smaller than the Hund's rule coupling. The two Cu$^{2+}$: $d^9$ (S=1/2) ions have one hole in the minority-spin $d_{x^2-y^2}$-like band with the highest occupied $d$ states showing predominantly $d_{z^2}$ character (with $z$ being set along the long Cu-O bond for each Cu). The in plane AFM coupling reduces the admixture between $d_{z^2}$ and $d_{x^2-y^2}$ states with respect to the non magnetic case. 

The slight differences in the band structure for up and down spin channels are due to the different environments of Cu1 
and Cu2. The magnetic moments for the Cu atoms are $\mu_{Cu1}$= -0.53 $\mu_B$ and $\mu_{Cu2}$= 0.52 $\mu_B$, reduced with respect to the nominal 1 $\mu_B$ value due to
hybridization. Hybridization between the Cu-$d$ and O-$p$ states along with the low site symmetries for the oxygens induces small moments on the oxygen ions:
$\mu_{O1}$= -0.01 $\mu_B$, $\mu_{O2}$=0.03 $\mu_B$, $\mu_{O3}$= 0.03 $\mu_B$ and $\mu_{O4}$= -0.05 $\mu_B$.
The sign of the Cu1 and O4 moments are the same, as are the Cu2 and O3 moments, driven by the fact that the short bonds are between
these ions.  The O2 moment is significantly larger than O1, consistent with the fact O1 forms a long bond with Cu1,
but O2 forms a medium bond with Cu2 (which in turn couples their two moments).

The spin density (Fig.~\ref{dos}) illustrates the two dimensional nature of the electronic structure. 
The shape of this density clearly reflects the d$_{x^2-y^2}$ orbital character around the Cu ions, with the lobes of the $d$-orbitals being directed along the short and medium Cu-O bonds (the medium one being out of plane for Cu2), giving rise to a direct overlap of the Cu-$d$ states with O-$p$ states of the neighboring oxygen ions.
Since the Cu-O-Cu bond angles are much larger than 90$^{\circ}$, one expects antiferromagnetic coupling, as we find, which is weaker than that in a typical cuprate which exhibits 180$^{\circ}$ bond angles.
Interestingly, the magnetism is essentially two-dimensional, and not chain-like, despite the fact that structurally, the material is composed
of S-M chains along the (1,0,1) planar directions, and S-L chains along the (1,0,-1) directions.
The coupling of states from one plane to the next proceeds via the Te ions, but this only has a minor effect on the spin density. Structural relaxations do not affect the electronic structure significantly.

Given the similarities of the predicted electronic structure of CuTeO$_4$ to typical cuprates as well as the prediction of a magnetic ground state in the stoichiometric
material, one might expect that a suitably doped phase would be a superconductor, which would be of significant interest given the very different Cu-O-Cu
bond angles in this material (with values more reminiscent of materials with triangular lattices like herbertsmithite \cite{herb}
than the 180$^\circ$ value found in an ideal cuprate).  In this context, the magnetic wave vector (2$Y$ in the notation of Fig.~1) couples
the Fermi surfaces to one another in Fig.~3 (it is a reciprocal lattice vector).  For the $d_{x^2-y^2}$ superconducting state in a cuprate, 
one would have line nodes on the lower two surfaces in Fig.~3, whereas a $d_{xy}$ state would have its nodes on the upper two.
Since a number of copper tellurates have
analogues where Te$^{6+}$ is replaced by Sb$^{5+}$, hole doping of this material is a realistic possibility.

This work was supported by the Center for Emergent
Superconductivity, an Energy Frontier Research Center funded by the
US DOE, Office of Science, under Award No.~DE-AC0298CH1088.
We acknowledge the computing resources provided on Blues and
Fusion, the high-performance computing clusters operated by the Laboratory
Computing Resource Center at Argonne National Laboratory.

\end{document}